\documentclass[10pt,letterpaper]{article}
\usepackage[top=0.85in,left=2.75in,footskip=0.75in,marginparwidth=2in]{geometry}

\usepackage[utf8]{inputenc}

\usepackage{cite}
\usepackage{amsmath}
\usepackage{nameref,hyperref}

\usepackage[right]{lineno}

\usepackage{microtype}
\DisableLigatures[f]{encoding = *, family = * }

\raggedright
\usepackage{changepage}

\usepackage[aboveskip=1pt,labelfont=bf,labelsep=period,singlelinecheck=off]{caption}

\makeatletter
\renewcommand{\@biblabel}[1]{\quad#1.}
\makeatother

\usepackage{lastpage,fancyhdr,graphicx}
\usepackage{epstopdf}
\fancyheadoffset[L]{2.25in}
\fancyfootoffset[L]{2.25in}

\usepackage{graphicx}
\usepackage{epstopdf}
\usepackage{latexsym}
\usepackage{amsmath,amssymb}
\usepackage{color}
\usepackage{xcolor}

\usepackage{algorithm}
\usepackage{algpseudocode}

\renewcommand{\algref}[1]{Algorithm~\ref{alg:#1}}
\newcommand{\lemref}[1]{Lemma~\ref{lem:#1}}
\renewcommand{\lineref}[1]{line~\ref{line:#1}}

\newcommand{\sfB}{{\mathsf B}}
\newcommand{\sfH}{{\mathsf H}}

\newcommand{\calB}{{\mathcal B}}
\newcommand{\calC}{{\mathcal C}}
\newcommand{\calH}{{\mathcal{H}}}
\newcommand{\calE}{{\mathcal{E}}}
\newcommand{\calI}{{\mathcal{I}}}
\newcommand{\calJ}{{\mathcal{J}}} %% added in R1

\newtheorem{lem}{Lemma}
\newtheorem{thm}{Theorem}
\newenvironment{proof}{\medskip
  \noindent{\scshape Proof:}}{\quad $\Box$\medskip}

\long\def\invis#1{}

\usepackage{graphicx}
\usepackage{epstopdf}
\usepackage{latexsym}
\usepackage{amsmath,amssymb}
\usepackage{color}
\usepackage{xcolor}

\usepackage{algorithm}
\usepackage{algpseudocode}
\usepackage{color}

\definecolor{Gray}{gray}{.25}

\usepackage{graphicx}

\usepackage{sidecap}

\usepackage{wrapfig}
\usepackage[pscoord]{eso-pic}
\usepackage[fulladjust]{marginnote}
\reversemarginpar

\begin{document}
\vspace*{0.35in}

% title goes here:
\begin{flushleft}
{\Large
    \textbf\newline{Systematic evaluation of the isolated effect of tissue environment on the transcriptome using a single-cell RNA-seq atlas dataset} }
\newline
% authors go here:
\\
Daigo Okada\textsuperscript{1*}, Jianshen Zhu \textsuperscript{2}, Kan Shota\textsuperscript{2}, Yuuki Nishimura\textsuperscript{2}, Kazuya Haraguchi\textsuperscript{2}

\bigskip
\bf{1} Center for Genomic Medicine, Graduate School of Medicine, Kyoto University, Kyoto 606-8507, Japan\\
\bf{2} Discrete Mathematics Laboratory, Applied Mathematics and Physics Course, Graduate School of Informatics, Kyoto University, Kyoto 606-8501, Japan\\
\bigskip
* Correspondence: dokada@genome.med.kyoto-u.ac.jp

\end{flushleft}

\section*{Abstract}

{\bf Background}: Understanding cellular diversity throughout the body is essential for elucidating the complex functions of biological systems. Recently, large-scale single-cell omics datasets, known as omics atlases, have become available. These atlases encompass data from diverse tissues and cell-types, providing insights into the landscape of cell-type-specific gene expression. However, the isolated effect of the tissue environment has not been thoroughly investigated. Evaluating this isolated effect is challenging due to statistical confounding with cell-type effects, arising from significant biases in the combinations of tissues and cell-types within the body. {\bf Results}: This study introduces a novel data analysis framework, named the Combinatorial Sub-dataset Extraction for Confounding Reduction (COSER), which addresses statistical confounding by using graph theory to enumerate appropriate sub-datasets. COSER enables the assessment of isolated effects of discrete variables in single cells. Applying COSER to the Tabula Muris Senis single-cell transcriptome atlas, we characterized the isolated impact of tissue environments. Our findings demonstrate that some of genes are markedly affected by the tissue environment, particularly in modulating intercellular diversity in immune responses and their age-related changes. {\bf Conclusion}: COSER provides a robust, general-purpose framework for evaluating the isolated effects of discrete variables from large-scale data mining. This approach reveals critical insights into the interplay between tissue environments and gene expression.

\subsection*{keyword}single cell RNA-seq; effect of tissue environment; graph theory; maximal biclique enumeration.

% now start line numbers (Arxivでは消す)
%\linenumbers

% the * after section prevents numbering
\section*{Background}
Understanding the cell diversity across the entire body and the underlying molecular mechanisms is essential for elucidating the complex functions of biological systems. Cells that differentiate from a fertilized egg develop into a wide variety of cell-types, existing in appropriate proportions within each tissue. Single-cell omics enables the acquisition of detailed omics information at the individual cell level and serves as a powerful tool for investigating this cell diversity \cite{okada2022cell,van2023applications}. Recently, large-scale single-cell omics datasets, referred to as omics atlases, have become available \cite{tabula2020single,he2020single,the2022tabula}. These datasets, which comprise data for a variety of tissues and cell-types, are valuable for comprehensive analyses of the effects of cellular features on gene expression using statistical models. However, while cell-type-related gene markers have been extensively studied in single-cell omics research \cite{Butler2018Integrating, Chao2021Identifying}, the impacts of tissue environments have received limited attention.

Although omics atlases encompassing multiple tissues are publicly available, evaluating the isolated effect of the tissue environment on gene expression presents considerable challenges. A primary issue in such evaluations is statistical confounding with cell-type effects due to substantial biases in the combinations of tissues and cell-types within the body. For example, while blood cells and fibroblasts are present in many tissues, certain cell-types are tissue-specific, such as hepatocytes in the liver or pancreatic cells in the pancreas. Furthermore, in tissues where cell collection is difficult, data may be limited to only a few cell-types. These factors contribute to statistical confounding between tissues and cell-types. Effectively addressing these challenges requires the development of novel data analysis techniques that account for the interrelationships among discrete variables.

Quantifying isolated tissue effects, though challenging, is fundamentally important for elucidating cellular diversity. When genes are expressed in the various cell-types that comprise a specific tissue but not in other tissues, it suggests that factors inherent to the tissue environment, such as cellular niches or secreted proteins, play a critical role in regulating gene expression. Conversely, when genes are expressed in multiple tissues within a given cell-type but not in different cell-types, internal cell-type effects, such as epigenetic status, are likely to be critical. Assessing the isolated effect of the tissue environment on gene expression is important for achieving a systems-level understanding of intercellular diversity within the transcriptome. Indeed, evidence from previous studies has highlighted the influence of the tissue environment on transcriptome data. For example, distinct age-related variations in gene expression have been observed in the same cell-types derived from different tissues \cite{kimmel2019murine}. Additionally, fibroblasts have been shown to exhibit heterogeneity among different tissues \cite{muhl2020single}. 

In this study, we introduce a novel data analysis framework, referred to here as Combinatorial Sub-dataset Extraction for Confounding Reduction (COSER). COSER enables the evaluation of the isolated effects of discrete variables in cells by overcoming statistical confounding by enumerating appropriate sub-datasets using graph theory. Application of this method to a large mouse scRNA-seq atlas dataset revealed the landscape of isolated tissue environment effects.

\section*{Result}
\subsection*{Visualization of bias in tissue and cell-type combinations in a single cell RNA-seq atlas}
This study utilized the Tabula Muris Senis (TMS) dataset, a large-scale publicly available mouse single-cell RNA-seq dataset \cite{tabula2020single}. 
The TMS dataset serves as a valuable resource for aging research \cite{zhang2021mouse, cheng2024data}, encompassing data from cells derived from 23 tissues collected from 30 mouse individuals across six age groups (1, 3, 18, 21, 24, and 30 months-old). The dataset includes a sex distribution of 19 males and 11 females. All cells in the dataset have been annotated with cell-types by the TMS project. We obtained log-transformed, pre-processed data from the TMS dataset, which comprises two subsets generated using distinct experimental methodologies: fluorescence-activated cell sorting (FACS) and droplet-based sequencing (Droplet). The FACS dataset contains expression data for 22,966 genes across 110,824 cells, while the Droplet dataset includes data for 20,138 genes across 245,389 cells. 

We visualized the bias in the tissue and cell-type combinations in each dataset (Figure 1). A bipartite graph was constructed wherein the edges represent existing combinations between tissues and cell-types in the dataset. The graphs included data from 23 tissues and 120 cell-types (FACS dataset) and 20 tissues and 123 cell-types (Droplet dataset). All edges of these graphs are shown in Supplementary File 1.
While some cell-types are present in multiple tissues, many are restricted to one or a few tissues, indicating a considerable bias in tissue and cell-type combinations. The bias inherent in the combinations of these discrete variables can lead to statistical confounding, thereby preventing accurate evaluation of individual effects.

Our research addresses the challenge of statistical confounding among discrete variables, including tissue and cell-type, when assessing the effects of the tissue environment on gene expression levels. Statistical confounding arises due to inherent biases in the combinations of these discrete variables, making it difficult to accurately evaluate individual effects. In theory, confounding can be resolved if all possible combinations of discrete variables were represented in the dataset. However, achieving this representation in real-world datasets is often impractical, necessitating the development of robust analytical methods to manage incomplete or biased combinations effectively.

\subsection*{Brief description of the COSER framework}
To address the issue of bias in the combination of discrete variables such as tissue and cell-type, we propose a novel data analysis framework called Combinatorial Sub-dataset Extraction for Confounding Reduction (COSER). By constructing a bipartite graph where the edges represent combinations of variables (e.g., Figure 1), within this graph, bicliques, i.e., subgraphs containing all possible combinations of connected variables, are identified (Figure 2A). These bicliques provide a robust foundation for statistical analysis by ensuring that variable combinations are comprehensively represented.

While identifying maximal bicliques in bipartite graphs is a well-established concept in graph theory, many biological datasets involve more than two discrete variables. To address this, we extended the maximal biclique enumeration problem to $k$-partite hypergraphs. For example, consider the combinations of three discrete variables of a single cell: sex, tissue, and cell-type. If all eight combinations exist in the dataset: male/liver/T-cell, male/liver/B-cell, male/spleen/T-cell, male/spleen/B-cell, female/liver/T-cell, female/liver/B-cell, female/spleen/T-cell, female/spleen/B-cell (Figure 2B), [[male, female], [liver, spleen], [T cell, B cell]] forms an extended biclique in a $k$-partite hypergraph (k = 3). The developed algorithm accepts as input a table representing combinations of discrete variables that exist in the dataset. The algorithm detects the maximal solutions with ensuring that each solution includes at least two distinct values for all variables represented in the dataset. Further details of this extension and the algorithm developed to enumerate all maximal solutions in $k$-partite hypergraphs are provided in the Method section.

Figure 2C shows an overview of the COSER framework as applied to scRNA-seq. First, all of the combinations of discrete variables in the dataset are listed. These combinations of discrete variables are represented as a $k$-partite hypergraph. The developed algorithm then enumerates subgraphs that contain all possible combinations, identifying them as solutions. For each solution, a sub-dataset is created that contains only the cells corresponding to the included combinations. Statistical analyses are performed independently on each sub-dataset, allowing for unbiased statistical evaluation of the effects of individual variables on cellular phenotypes, such as gene expression levels. By integrating the results of these independent statistical analyses, a consensus conclusion is reached, providing robust insights into the isolated effects of the variables.

As an example implementation, we applied the COSER framework to the bipartite graph of the FACS dataset (Figure 1). As a result, 31 maximal solutions were identified (Supplementary File 2). These solutions represent suitable units of analysis for evaluating the impacts of the tissue environment or cell-type effects on cell phenotypes within the dataset. For example, Example 1 in Figure 2D contains four adipose sub-tissues (i.e., brown adipose tissue (BAT), gonadal adipose tissue (GAT), mesenteric adipose tissue (MAT), and subcutaneous adipose tissue (SCAT)), and represents the solution with the largest number of edges. In contrast, Example 2 in Figure 2D features another maximal solution and includes a more anatomically diverse set of tissues. These maximal solutions form bipartite cliques, which means that all of their combinations are included in the original dataset. Researchers can explore these enumerated maximal solutions to identify solutions that align with their research questions.

\begin{figure}[ht]
\centering
\includegraphics[width=0.9\textwidth]{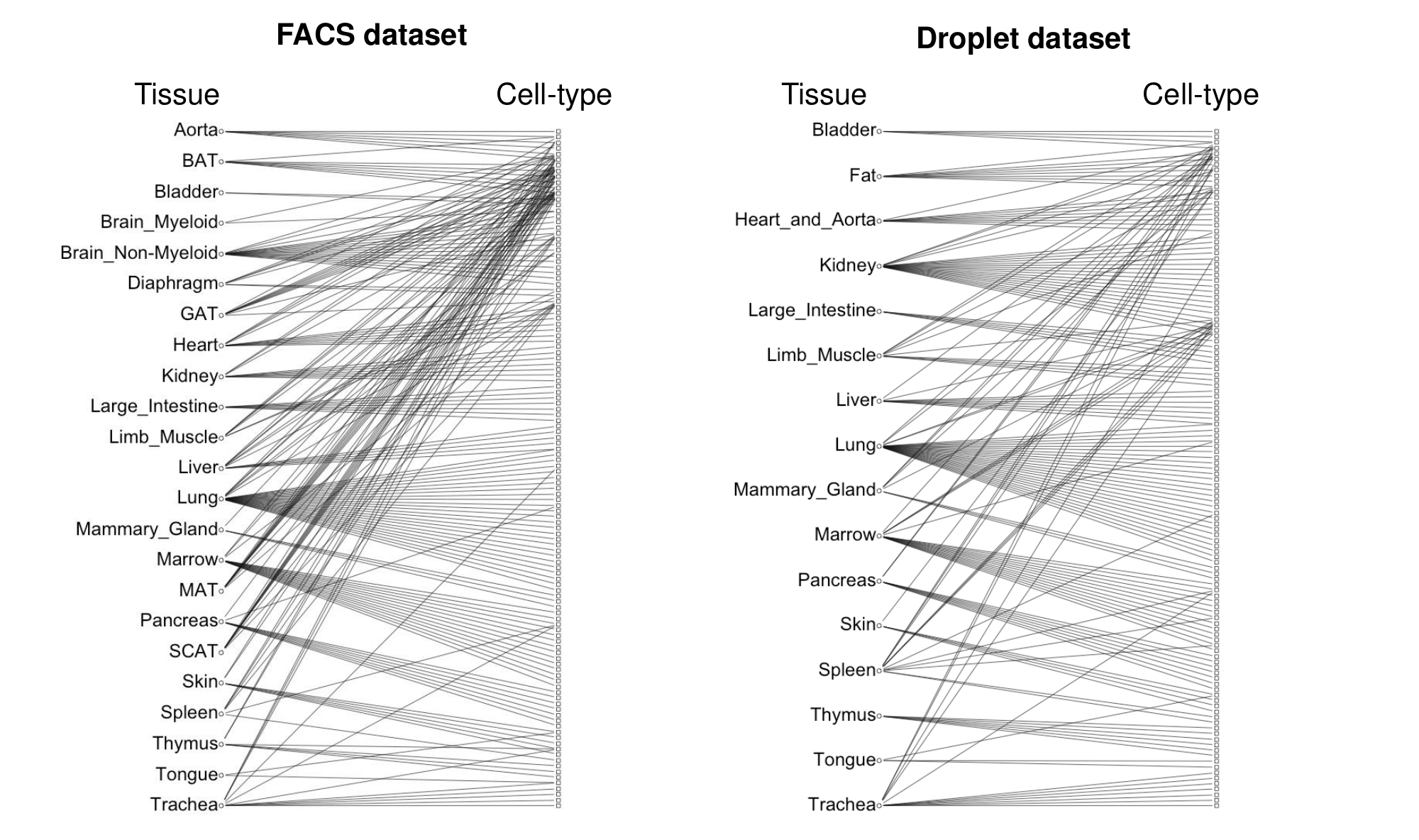}
\caption{Bipartite graphs of tissue and cell-type combinations in the TMS dataset. Biases in the combinations of tissues and cell-types are shown, highlighting the imbalance in their representation. }
\end{figure}

\begin{figure}[ht]
\centering
\includegraphics[width=0.9\textwidth]{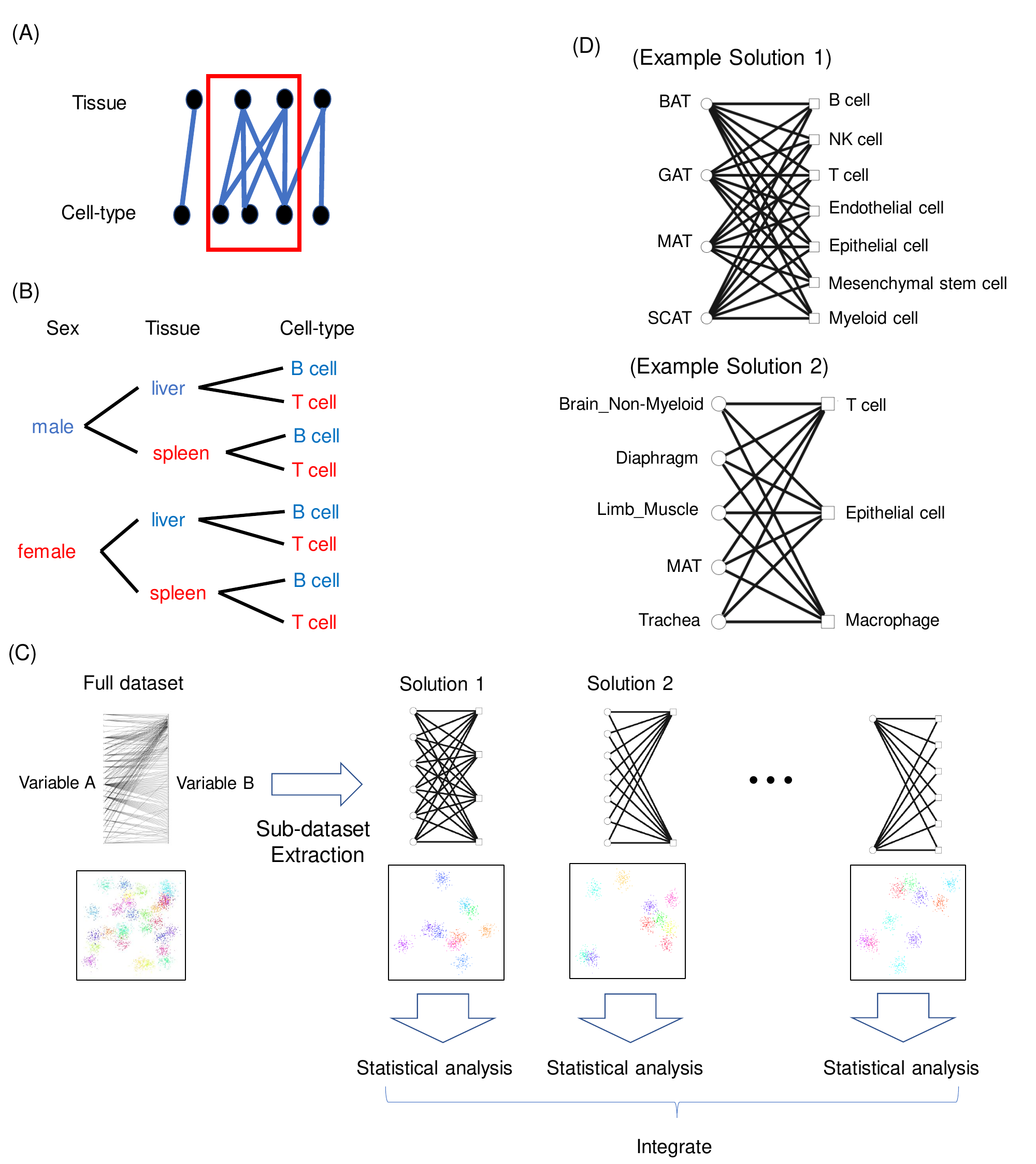}
\caption{ Extension of the maximal biclique enumeration problem to $k$-partite hypergraphs and the COSER framework. (A) Illustration of a maximal biclique. (B) An example of extending bicliques to $k$-partite hypergraphs, where the solution [[male, female], [liver, spleen], [T cell, B cell]] ensures the presence of all eight combinations shown in the tree diagram within the dataset.
(C) Graphical overview of the COSER framework. These combinations of discrete variables in dataset are represented as a $k$-partite hypergraph. The subgraphs that contain all possible combinations are identified as solutions. For each solution, a sub-dataset is created that contains only the cells corresponding to the included combinations. Independent statistical analyses are conducted on sub-datasets, and their results are integrated to derive a consensus, ensuring robust insights into the variables' isolated effects.
(D) Examples of maximal bicliques in the bipartite graph of FACS dataset shown in Figure 1.}
\end{figure}

\subsection*{Quantitative assessment of the isolated effect of the tissue environment}

We used COSER to examine the isolated effect of the tissue environment on single-cell transcriptome profiles. This analysis specifically targeted cells from mice aged three months, focusing on the combinations of individuals, tissues, and cell-types. Only combinations containing more than ten cells were selected for downstream analysis. We applied COSER to a three-column table comprising individual, tissue, and cell-type combinations to enumerate the maximal solutions. We then quantified the tissue effect on gene expression levels using a generalized liner model (GLM) for each sub-dataset. In this model, gene expression values were treated as the objective variable, while individual, tissue, and cell-type were used as explanatory variables.

In the FACS dataset, 24 maximal solutions for individual $\times$ tissue $\times$ cell-type combinations were identified (Figure 3A, Supplementary File 3). Figure 3B shows a QQ plot of the P-values obtained for the tissue and cell-type effects in each solution, demonstrating that not only cell-type but also tissue has an isolated effect on gene expression. The P-values for each gene in all sub-datasets are listed in Supplementary File 4. Genes exhibiting a significant tissue effect in more than 12 of the 24 sub-datasets (FDR $<$ 0.05) were identified as tissue environment-susceptible genes. A total of 253 such genes were identified. Enrichment analysis of Gene Ontology (GO) Biological Processes for these genes revealed 135 significantly enriched GO terms within this gene set (FDR $<$ 0.05) (Supplementary File 5). The ten GO terms with the highest enrichment scores are shown in Figure 3C. The most significantly enriched biological process was GO:0035455 (response to interferon-alpha). Other biological processes related to immune responses, such as GO:0097028 (dendritic cell differentiation), GO:0035456 (response to interferon-beta), and GO:0070670 (response to interleukin-4), were also highly represented. Additionally, a fundamental cellular function, GO:0002181 (cytoplasmic translation), also appeared among the enriched terms.

In the Droplet dataset, a single maximal solution comprising two tissues (limb muscle and mammary gland) was identified (Figure 3A). As with the FACS dataset, isolated tissue effects were observed. A statistically significant contribution from the tissue environment was detected in 3,581 genes (FDR $<$ 0.05). The P-values for all genes are shown in Supplementary File 6. Figure 3D shows a QQ plot of the P-values obtained for tissue effects. 
Among these genes, 264 GO terms were significantly enriched (FDR $<$ 0.05) (Supplementary File 7). The 10 GO terms with the highest enrichment scores are shown in Figure 3D. Despite this result being based on a comparison between limb muscle and mammary gland, processes related to immune response and cytoplasmic translation were prominently represented, which is consistent with the FACS dataset.

\begin{figure}[ht]
\centering
\includegraphics[width=0.9\textwidth]{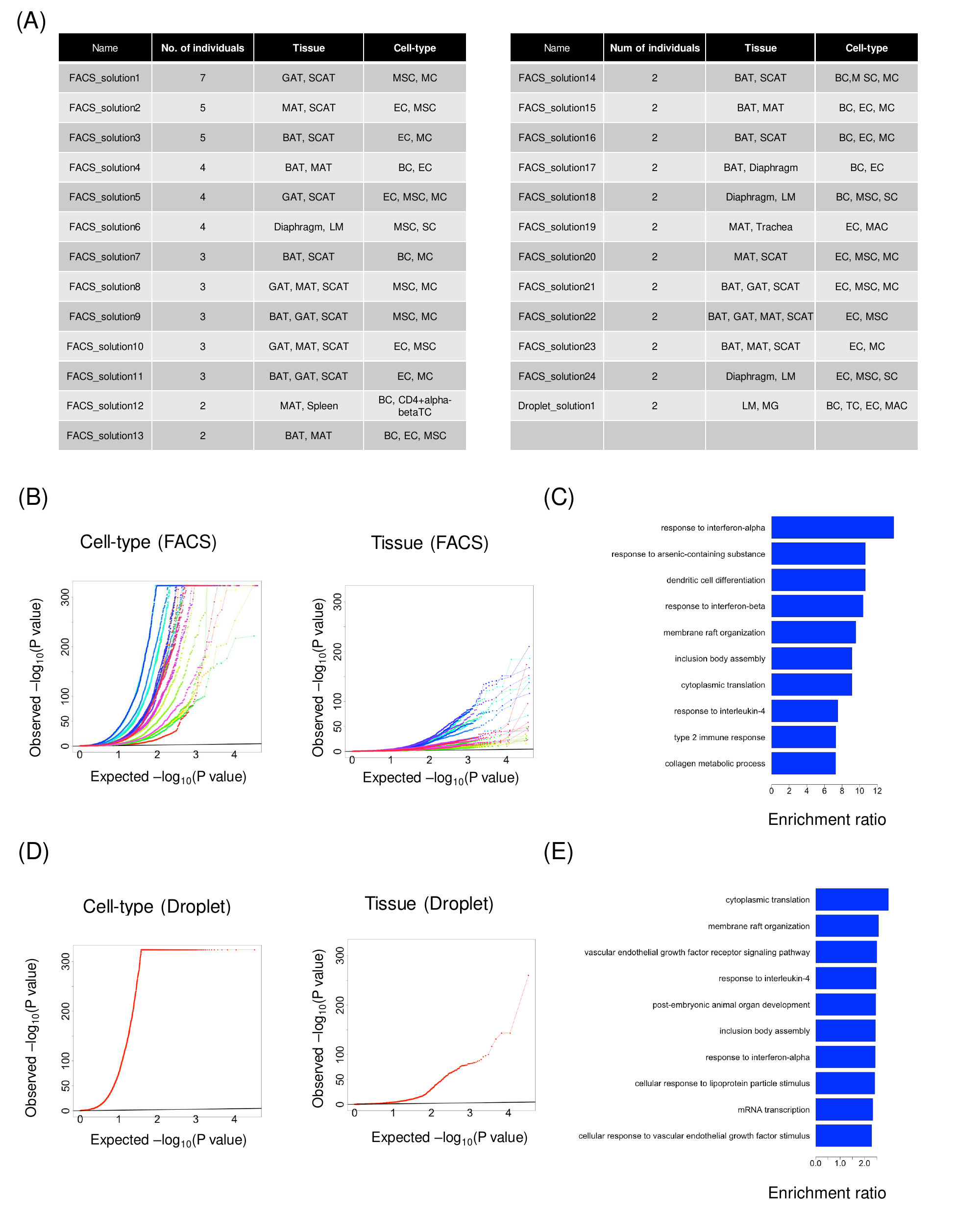}
\caption{Isolated effects of the tissue environment observed throughout the body. (A) All maximal solutions for the combinations of individuals, tissues, and cell-types. 
BAT: brown adipose tissue, GAT: gonadal adipose tissue, MAT: mesenteric adipose tissue, SCAT: subcutaneous adipose tissue, LM: limb muscle, MG: mammary gland, MSC: mesenchymal stem cell/mesenchymal stem cell of adipose, MC: myeloid cell, BC: B cell, TC: T cell, SC: skeletal muscle satellite cell, MAC: macrophage. 
(B) QQ plot showing P-values for the effect of tissue and cell-type in 24 sub-datasets from the FACS dataset. Each line corresponds to a sub-dataset. Zero P-values were replaced with the minimum non-zero P-value before log transformation.
(C) GO terms with the top ten enrichment scores for genes affected by the tissue environment in the FACS dataset. (D) QQ plot showing P-values for the effect of the tissue and cell-type in one sub-dataset from the FACS dataset. (E) GO terms with the top ten enrichment scores for genes affected by the tissue environment in the Droplet dataset.
}
\end{figure}

By integrating the results of statistical analyses from the sub-datasets corresponding to each solution, it is possible to compare isolated tissue effects among tissues. To achieve this, a directed graph was constructed with tissues represented as nodes, based on the order of the coefficients of the tissue effects obtained from all 24 analyses in the FACS dataset. If this directed graph forms a directed acyclic graph (DAG), then the tissue effects are considered to have a partial ordering structure. Using the constructed DAG, a consistent order of tissue effects on gene expression was obtained. Among the identified tissue environment-susceptible genes, a DAG was successfully constructed for 54 genes (Supplementary File 8). Notably, four transcription factors were identified within these genes ({\it Fosb}, {\it Klf4}, {\it Tbx15}, {\it Wt1}).

Figure 4 shows the DAGs constructed for the four transcription factor genes. These DAGs provide valuable insights into the differences in gene expression among adipose sub-tissues. For example, the effect of the tissue environment on {\it Tbx15} gene expression follows the order: SCAT $>$ BAT $>$ GAT $>$ MAT. It has been reported that {\it Tbx15} is highly expressed in brown adipose tissue and essential for differentiating brown adipocytes \cite{Gburcik2012An}. The findings of this study suggest that the adipose tissue environment may affect the adipocyte differentiation through the regulation of {\it Tbx15} expression. By integrating the results of statistical model analyses for each sub-dataset into a graph where nodes represent tissues, this approach enables a systematic evaluation of the relative magnitude of tissue environment effects on gene expression.

\begin{figure}[ht]
\centering
\includegraphics[width=0.9\textwidth]{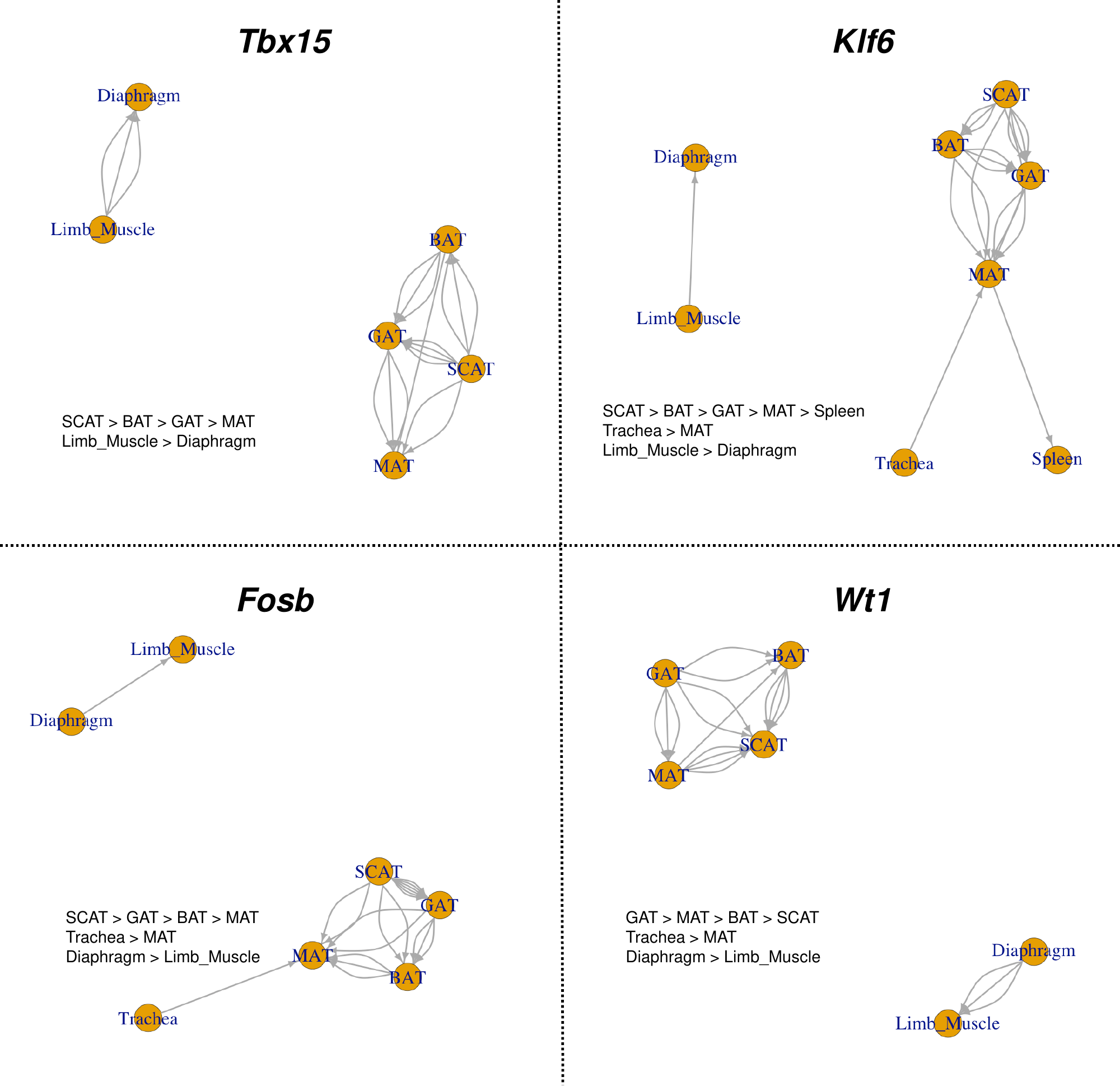}
\caption{ DAGs representing the relative magnitude of the isolated effect of tissue environments on the expression of four transcription factors ({\it Fosb}, {\it Klf4}, {\it Tbx15}, {\it Wt1}). The inequalities of the isolated tissue effects derived from the graphs are also included.
}
\end{figure}

\subsection*{Detection of divergent aging patterns between different tissues}

We investigated transcriptomic changes associated with aging in different tissue environments. Previous studies have demonstrated that age-related changes in gene expression can be specific to both tissue and cell-type \cite{Stegeman2017Transcriptional, zhang2021mouse, Ham2020Advances}. In addition, data-mining analyses of the TMS dataset have identified genes that exhibit age-related increases in expression in specific sex-tissue-cell-type combinations while showing decreases in others (or vice versa) \cite{okada2024opposite}. Understanding the underlying mechanisms driving these differences in age-related changes is crucial for advancing our knowledge of tissue-specific aging processes.

Using COSER, we investigated the occurrence of opposite aging effects within different tissue environments. For both the FACS and Droplet datasets, we focused on different combinations of sex, tissue, cell-type, and age. Donor age was defined as ``Young'' for three-month-old cells, ``Old'' for cells aged $\geq$ 18 months, and cells from one-month-old donors were excluded. Only combinations harboring more than 25 cells were selected for further analysis. When we applied COSER to this four-column table to identify sub-datasets, four maximal solutions were obtained using the FACS dataset while none were obtained using the Droplet dataset (Figure 5A).

We focused on two solutions from the FACS dataset, each containing at least three cell-types (FACS\_solution1 and FACS\_solution3). The sub-dataset corresponding to FACS\_solution1 facilitates a comparative analysis of gene expression between GAT and SCAT. In contrast, the sub-dataset corresponding to FACS\_solution3 facilitates the exploration of differences between MAT and SCAT.

We performed GLM analysis on cells from each tissue in a sub-dataset, and calculated the regression coefficients and P-values for the ``Young''  category across all genes in each tissue. We identified genes where the sign of the ``Young''  coefficient was opposite between tissues and both were statistically significant (FDR $<$ 0.05).
Scatter plots of the regression coefficients for the ``Young'' category in the two tissues are shown in Figures 5B and 5C. A total of 14 genes in the BAT vs. SCAT comparison and 119 genes in the MAT vs. SCAT comparison exhibited different directions of significant age-related gene expression change. Detailed results, including regression coefficients for BAT vs. SCAT and MAT vs. SCAT, are shown in Supplementary Table 9 and Supplementary Table 10, respectively.

The findings showed that opposite aging effects are present even among different adipose sub-tissues. Specifically, four GO terms were significantly associated with genes exhibiting age-related expression changes in opposite directions between BAT and SCAT. These terms include GO:0042742 (defense response to bacterium), GO:0042107 (cytokine metabolic process), GO:0002237 (response to molecule of bacterial origin), and GO:0002526 (acute inflammatory response), as shown in Figure 5D and Supplementary Table 11. These results suggest that gene expression related to immunity and inflammation changes with age in opposing directions, depending on the tissue environment. Such findings highlight the importance of accounting for the influence of the tissue environment when studying age-related changes in immunity and inflammation. In the case of MAT vs. SCAT, no GO terms were significantly enriched.

\begin{figure}[ht]
\centering
\includegraphics[width=0.9\textwidth]{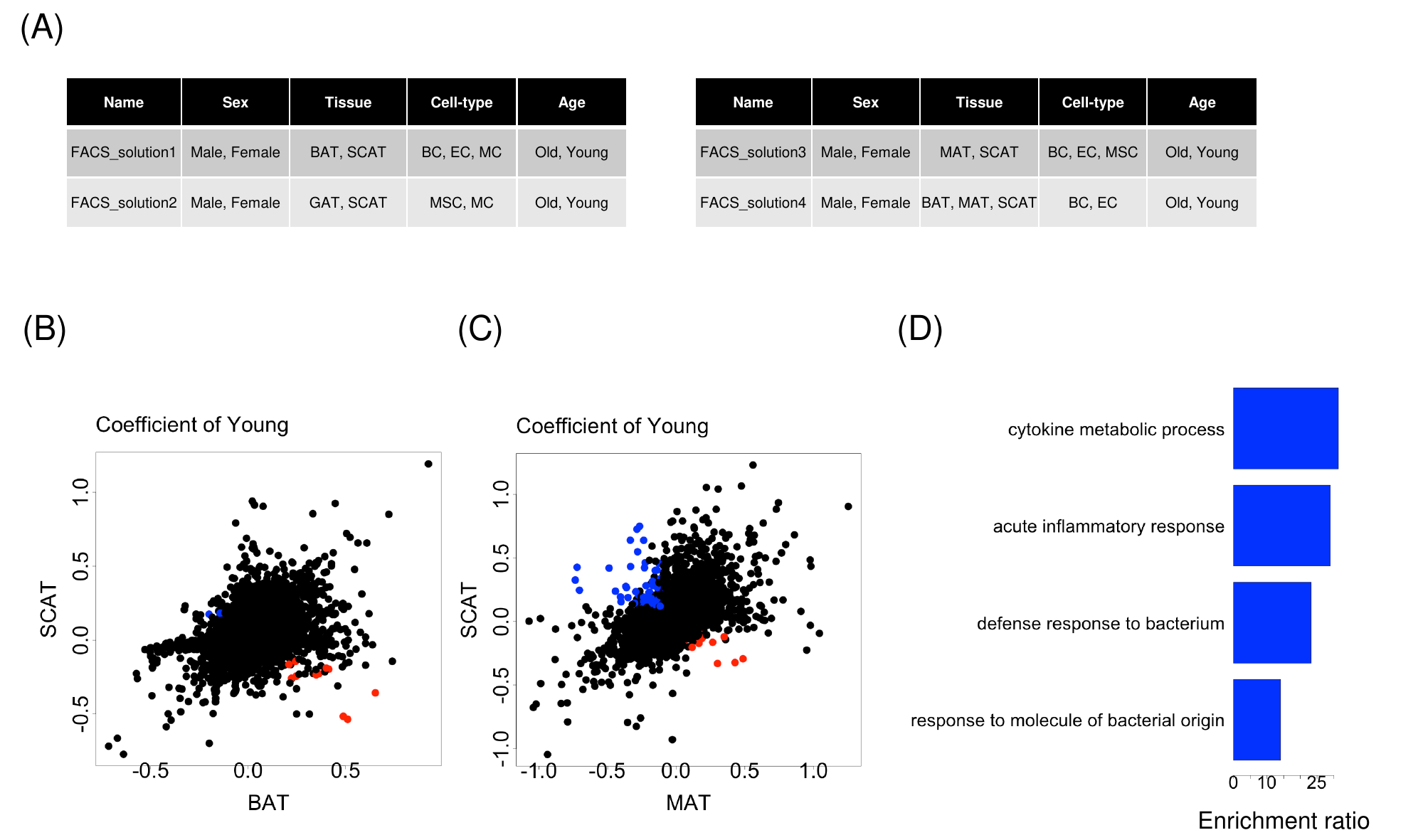}
\caption{Comparison of age-related changes between different tissue environments. (A) The four maximal solutions found for each combination of tissue, cell-type, sex, and age. (B) Scatter plot of regression coefficient of the ``Young'' to gene expression level. Genes exhibiting the opposite effect of aging are colored red (X-axis is positive) or blue (Y-axis is positive). (C) Scatter plot of the ``Young'' regression coefficient versus the gene expression level (MAT vs. SCAT). (D) GO terms with the top ten enrichment scores for genes exhibiting the opposite effect of aging (BAT vs. SCAT). 
}
\end{figure}

\section*{Discussion}
In this study, we showed that a substantial number of genes are affected by the tissue environment. Specifically, various immune response pathways were associated with genes exhibiting differential expression affected by tissue environments. In addition, biological processes related to translation showed a strong association with the tissue environment. These findings highlight the importance of considering the tissue environment when analyzing cellular gene expression patterns. This investigation represents a pioneering effort to systematically evaluate the isolated effects of the tissue environment through data mining of a large-scale scRNA-seq atlas.

Aging analysis suggested that age-related changes in gene expression are influenced by the tissue environment. Specifically, genes exhibiting age-related changes were predominantly enriched in biological functions associated with immune responses. This decline in immune function with age, referred to as immunosenescence, is an important aspect of the aging process in individuals \cite{Aw2007Immunosenescence, Wang2022Immunosenescence}. While the tissue environment typically regulates the cellular states involved in the immune responses, aging may disrupt this regulatory process.

There are several limitations to this study. First, the tissues were not evenly distributed in the detected solutions. For example, although the FACS dataset contains 23 tissues, only 8 were included in any of the 24 identified solutions. On the other hand, certain tissues, such as adipose tissue, were repeatedly detected in multiple solutions, allowing for detailed examination but preventing a comprehensive evaluation of all tissues in the body. In addition, when comparing the FACS dataset and the Droplet dataset from the same project, there was a significant difference in their applicability for evaluating tissue effects. While multiple solutions were obtained for the FACS dataset, the results obtained for the Droplet dataset were more limited. This discrepancy arises because the TMS dataset was not specifically designed to quantify tissue effects. Instead, the findings highlight the potential application of COSER to future research in experimental planning and statistical modeling for omics atlases.

In this study, we developed novel data analysis framework, COSER, to enumerate suitable sub-datasets based on the combinations of discrete variables within a dataset. While selecting sub-datasets is an effective strategy for mitigating statistical confounding, manual extraction of suitable analysis units can be challenging. COSER addresses this issue by systematically selecting sub-datasets through the extension of the maximal bipartite clique enumeration problem to a $k$-partite hypergraph. From a future perspective, applying this approach to datasets from various omics layers has the potential to uncover the overall diversity and functional landscape at the cellular level, thereby contributing to advances in the life sciences. Although this study primarily focused on the effect of the tissue environment in a single scRNA-seq atlas dataset, the COSER framework could be applied to any dataset containing multiple discrete variables.

\section*{Method}
\subsection*{Acquisition of data}
The pre-processed scRNA-seq data of TMS dataset were obtained using the ``TabulaMurisSenisData'' package in R \cite{TabulaMurisSenisData}. The list of mouse transcription factors was obtained from AnimalTFDB \cite{shen2023animaltfdb}.

\subsection*{Algorithm for sub-dataset extraction by extending the maximal biclique enumeration problem to $k$-partite hypergraphs}
%\section{Method}
We model a table on experimental results (i.e., a dataset) by a hypergraph. 
A {\em hypergraph} $\sfH=(V,\calE)$ is a pair
of a  set $V$ of {\em vertices}
and a set $\calE$ of {\em hyperedges}, where $\calE\subseteq 2^V$.
For an integer $k \ge 2$,
$\sfH$ is \emph{$k$-partite}
if there is a partition
$V = V_1 \cup V_2 \cup \cdots \cup V_k$
such that, for every hyperedge $H \in \calE$,
$|H \cap V_i| = 1$, $i = 1, 2, \dots, k$ holds.
%%%%% R1 %%%%%
A conventional bipartite graph is a 2-partite hypergraph in this terminology.  
For a subset $S\subseteq V$,
we denote $S_i:=S\cap V_i$, $i=1,2,\dots,k$.
For a subset $S$ such that all of $S_1,S_2,\dots,S_k$ are non-empty
and $q\in\{1,2,\dots,k\}$, 
we define $\Pi_q(S)\triangleq\{\{v_1,v_2,\dots,v_q\}\mid v_i\in S_i,\ i=1,2,\dots,q\}$. In other words,
$\Pi_q(S)$ is the set of all combinations of vertices
that are taken from $S_1,S_2,\dots,S_q$ one by one, respectively. 
%%%%%%%%%%%%%%

%%For subsets $S_1,S_2,\dots,S_q\subseteq V$ of vertices
%%that are pairwise disjoint,
%%we define $\Pi(S_1,S_2,\dots,S_q)\triangleq\{\{v_1,v_2,\dots,v_q\}\mid v_i\in S_i,\ i=1,2,\dots,q\}$. 

Let $\sfH=(V,\calE)$ be a $k$-partite hypergraph 
and $\theta_1,\theta_2,\dots,\theta_k$ be positive integers. 
%%%
We call a tuple $(\sfH;\theta_1,\theta_2,\dots,\theta_k)$
an {\em instance}. 
We call $S\subseteq V$ a {\em solution to the instance}
if
\begin{itemize}
\item $|S_i|\ge\theta_i$ holds for all $i=1,2,\dots,k$; and
\item $\Pi_k(S)\subseteq\calE$ holds. %where $\prod$ denotes the Cartesian product. 
\end{itemize}
A solution $S$ is {\em maximal} if there is no solution
that is a proper superset of $S$.
It is easy to see that, when $k=2$ and $\theta_1=\theta_2=1$,  
a solution corresponds to a biclique in a bipartite graph.
Parameter $\theta_i$, $i=1,2,\dots,k$
is determined by users and
represents a threshold on the number of entries in $V_i$ that is regarded as significant. 

In our context,
a table like Figure~5A %%on experimental results
can be represented by a $k$-partite hypergraph such that 
each vertex corresponds to an entry in the table
(e.g., ``Female'', ``BAT", ``BC'', ``Old'');
each subset $V_i$, $i=1,2,\dots,k$ in the partition
corresponds to a column of the table
(e.g., ``Sex'', ``Tissue'', ``Cell-type'', ``Age"); and
each hyperedge $H\in\calE$ corresponds to
a row of the table. 
A solution $S$ to the instance $(\sfH;\theta_1,\theta_2,\dots,\theta_k)$
corresponds to
a set of rows in the dataset
such that all possible
$|S_1|\times|S_2|\times\dots\times|S_k|$
combinations of entries appear,
where $|S_i|\ge\theta_i$, $i=1,2,\dots,k$ holds. 
%
%In our experiments, we will construct $\sfH$ from a given dataset with $k=3$
%as above, 
%and set $\theta_1=\theta_2=\theta_3:=2$.
In our experiments, we will construct $\sfH$ from a given dataset as above,
and set $\theta_1=\theta_2=\cdots=\theta_k:=2$.
%%%%%%%%%%%%%%

We consider the problem of enumerating all maximal solutions.
%%%%% R1 %%%%%
Let $n:=|V|$, $m:=|\calE|$, and $N$ denote the number of all maximal solutions.
It is inevitable that the enumeration task takes $\Omega(N)$ time,
and $N$ can be up to an exponential number with respect to
$n$ and $m$. 
For example, when $k=2$, there exist $2^{\min\{|V_1|,|V_2|\}}$
maximal solutions in crown graphs.
%%%%%
A natural question is to ask whether or not
we can enumerate all maximal solutions in polynomial time with respect to
$n$, $m$ and $N$ (i.e., output-polynomial time~\cite{JYP.1989}).
%%%%%
Unfortunately, it is hopeless to obtain such an algorithm 
since it is NP-complete to decide
whether there exists a solution
even for the case of $k=2$ and $\theta_1=\theta_2$~\cite{garey1979computers}.
This indicates that there is no polynomial-time algorithm
to find a maximal solution unless P$=$NP.

%%%
Based on the hardness of the problem,
we decide to focus on developing
an algorithm that completes the enumeration task
for our datasets in practical time
while the time complexity bound is trivial $O^\ast(2^n)$,
where $O^\ast(\cdot)$ ignores polynomial factors. 
For space complexity, the algorithm
uses $O^\ast(2^n)$ space to store all candidates of maximal solutions. 
However, %as we will see in the next sections,
the algorithm is efficient enough for our datasets. 
 
%%%%%%%%%%%%%%%%%%%%%%%%%%%%%%%%%%%%%%%%%%%%%%%%%%%%%%%%%%%%

%%%%% R1 %%%%%
Let us introduce notations for preparation.
%%%%%%%%%%%%%%
For a vertex $v\in V$,
we denote by $\calE(v)$ the set of all hyperedges in $\calE$
that contain $v$.
We define $\hat{\calE}(v)$ to be the family
of all vertex subsets that are obtained by deleting $v$
from a hyperedge in $\calE(v)$; i.e., 
$\hat{\calE}(v)\triangleq\{H\setminus\{v\}\mid H\in\calE(v)\}$.
For $i\in\{1,2,\dots,k\}$, let $U\subseteq V_i$.
We define $\hat{\calE}(U):=\bigcap_{v\in U}\hat{\calE}(v)$. 
%%%%%%%%%%
For any subset $F\in\hat{\calE}(U)$ and vertex $v\in U$,
the union $F\cup\{v\}$ is a hyperedge in $\calE$.
%%%%%%%%%%
For a $k$-partite hypergraph $\sfH$,
let us define an auxiliary bipartite graph 
$\sfB_\sfH=(V_k\cup W_{\sfH},E_\sfH)$ such that
$W_{\sfH}:=\bigcup_{v\in V_k}\hat{\calE}(v)$
%%and $E_{\sfH}:=\{(v,\hat{H})\in V_k\times W_{\sfH}\mid \{v\}\cup\hat{H}\in\calE\}$.
and $E_{\sfH}:=\{(v,\hat{H})\mid v\in V_k,\hat{H}\in W_{\sfH}, \{v\}\cup\hat{H}\in\calE\}$. 
Note that each node
$\hat{H}\in W_{\sfH}$ is a subset of $V_1\cup V_2\cup\dots\cup V_{k-1}$. 

%%%%% R1 %%%%%
In order to avoid redundant search, we restrict the search space
by using two necessary conditions
that should be satisfied by any maximal solution.
Let $\calI=(\sfH=(V,\calE);\theta_1,\theta_2,\dots,\theta_k)$. 
First, if $S$ is a maximal solution to $\calI$,
then there is a maximal solution $B$ to the auxiliary instance
$\calJ=(\sfB_\sfH;\theta_k,1)$ such that $S_k=B\cap V_k$. 
This indicates that
$S_k$ of maximal solutions $S$ to $\calI$
are within maximal solutions to $\calJ$,
where the latter solutions can be efficiently computed by existing
biclique enumeration algorithms. %%e.g., \cite{Zhang:2014aa}. 
%%%
Second, if $S$ is a maximal solution to $\calI$,
then $S\setminus S_k=S_1\cup S_2\cup\dots\cup S_{k-1}$ must be
a maximal solution
in the reduced instance $\calI'=(\sfH';\theta_1,\theta_2,\dots,\theta_{k-1})$, 
where $\sfH':=(V_1\cup V_2\cup\dots\cup V_{k-1},\hat{\calE}(S_k))$.
Using this property,
we generate the candidates of $S_k,S_{k-1},\dots,S_1$ recursively
and maintain $S_k\cup S_{k-1}\cup\dots\cup S_1$
as a candidate of a maximal solution to $\calI$. 
After generating all candidates,
we output those which are inclusion-wise maximal. 

The two necessary conditions are summarized as Lemmas~\ref{lem:bipartite}
and \ref{lem:maximal} as follows. 
%%%%%%%%%%%%%%

\begin{lem}
  \label{lem:bipartite}
  For an instance $\calI=(\sfH=(V,\calE);\theta_1,\theta_2,\dots,\theta_k)$,
  if $S\subseteq V$ is a maximal solution to $\calI$,
  then $S_k\cup\hat{\calE}(S_k)$ is a maximal solution to $\calJ=(\sfB_\sfH;\theta_k,1)$. 
\end{lem}
\begin{proof}
  We see that $\hat{\calE}(S_k)=(\bigcap_{v\in S_k}\hat{\calE}(v))\subseteq(\bigcup_{v\in V_k}\hat{\calE}(v))=W_{\sfH}$.
  For every $u\in S_k$ and $\hat{H}\in\hat{\calE}(S_k)$,
  it holds that $(u,\hat{H})\in E_\sfH$
  since $\hat{H}\in\hat{\calE}(S_k)=\bigcap_{v\in S_k}\hat{\calE}(v)\subseteq\hat{\calE}(u)$,
  indicating that $\{u\}\cup\hat{H}\in\calE$.
  Then $S_k\cup\hat{\calE}(S_k)$ is a solution to $\calJ$,
  where the maximality is obvious. 
\end{proof}

%% The following \lemref{maximal}(ii)
%% shows a necessary condition of a maximal solution;
%% a maximal solution $S$ to $(\sfH;\theta_1,\theta_2,\dots,\theta_k)$
%% should contain a maximal solution to
%% $(\sfH';\theta_1,\theta_2,\dots,\theta_{k-1})$ as a subset, 
%% where $\sfH'$ is a $(k-1)$-partite hypergraph
%% that can be obtained from $\sfH$ and $S$. 

\begin{lem}
  \label{lem:maximal}
  Suppose that we are given an instance
  $\calI=(\sfH=(V,\calE);\theta_1,\theta_2,\dots,\theta_k)$. 
  %%that consists of a $k$-partite hypergraph $\sfH=(V,\calE)$
  %%with a vertex partition $V=V_1\cup V_2\cup\dots\cup V_k$
  %%and positive integers $\theta_1,\theta_2,\dots,\theta_k$. 
  %%
  Let $S$ be a solution to $\calI$
  and $\sfH':=(V_1\cup V_2\cup\dots\cup V_{k-1},\hat{\calE}(S_k))$. 
  %%$i=1,2,\dots,k$.
  %%
  {\rm (i)} The hypergraph $\sfH'$
  is $(k-1)$-partite. 
  {\rm (ii)}
  If $S$ is a maximal solution to $\calI$, 
  then $S\setminus S_k$ is a maximal solution
  to $\calI':=(\sfH'; \theta_1,\theta_2,\dots,\theta_{k-1})$. 
\end{lem}
\begin{proof}
  (i) We see that
  $\hat{\calE}(S_k)=\bigcap_{v\in S_k}\hat{\calE}(v)$
  by the definition,
  where each $\hat{H}\in\hat{\calE}(v)$ satisfies
  $|\hat{H}\cap V_j|=1$ for $j=1,2,\dots,k-1$
  since $\sfH$ is $k$-partite.
  This shows that
  $\sfH'$ is $(k-1)$-partite.
  %%where $V\hat{\calE}(S_i)]\subseteq V\setminus V_i$ holds.

  \noindent
  (ii) By (i), $\sfH'$ is $(k-1)$-partite.
  Obviously $|S_i|\ge\theta_i$ holds for $i=1,2,\dots,k-1$. 
  Let $H:=\{v_1,v_2,\dots,v_{k-1}\}\in\Pi_{k-1}(S)$. %%\prod_{i=1}^{k-1}S_i$.
  For every $v_k\in S_k$,
  we have $H\cup\{v_k\}\in\calE(v_k)\subseteq\calE$
  since $S$ is a solution to $\calI$.
  Then $H\in\bigcap_{v_k\in S_k}\hat{\calE}(v_k)=\hat{\calE}(S_k)$ holds, where we see that $S\setminus S_k$ is a solution to $\calI'$.
  If $S\setminus S_k$ is not maximal, 
  then there would be a solution $S^+$ to $\calI'$
  such that $S^+\supsetneq S\setminus S_k$.
  For $u\in S^+\setminus (S\setminus S_k)$,
  it is easy to see that $S\cup\{u\}$ is a solution
  to $\calI$, contradicting the maximality of $S$. 
\end{proof}

%%%%%%%%%%%%%%%%%%%%%%%%%%%%%%%%%%%%%%%%%%%%%%%%%%%%%%%%%%%%
Now we are ready to present an algorithm
to enumerate all maximal solutions.
The algorithm is summarized as \textsc{EnumMaxSol}
in \algref{enummaxsol}. 
%%To obtain maximal solutions to a 2-partite instance
%%in \lineref{bipartite}, we can make use of
%%an existing algorithm; e.g., \cite{Zhang:2014aa}. 

\begin{algorithm}[t!]
\caption{An algorithm \textsc{EnumMaxSol}$(\calI)$ to output all maximal solutions to a given instance $\calI$}
\label{alg:enummaxsol}
\begin{algorithmic}[1]
  \Require An instance $\calI=(\sfH;\theta_1,\theta_2,\dots,\theta_k)$
  that consists of a $k$-partite hypergraph $\sfH=(V,\calE)$
  and positive integers $\theta_1,\theta_2,\dots,\theta_k$
  \Ensure All maximal solutions to $\calI$
  \State $\calC:=\emptyset$;
  \If {$k=1$ and $|\calE|\ge\theta_1$}
  \State $S:=$~the set of vertices that belong to hyperedges in $\calE$;
  %% \State $\calC:=\calC\cup\{S\}$
  \State $\calC:=\{S\}$
  \Else
  \State $\calB:=$~the set of maximal solutions to $\calJ=(\sfB_\sfH;\theta_k,1)$;
  \label{line:bipartite}
  \ForAll{$B \in \mathcal{B}$}
  \State $K:=B\cap V_k$;\label{line:K}
  \State $\calI':=((V_1\cup V_2\cup\dots\cup V_{k-1},\hat{\calE}(K));\theta_1,\dots,\theta_{k-1})$;
  \label{line:subproblem}
  \State $\hat{\calH}:=$~\textsc{EnumMaxSol}$(\calI')$;
  \label{line:recursive}
  \ForAll {$\hat{H}\in\hat{\calH}$}
  \State $S:=K\cup\hat{H}$;
  \label{line:union}
  \If{there is no $S^+\in\calC$ such that $S^+\supseteq S$}
  \label{line:if}
  \State $\mathcal{C} := \mathcal{C} \cup \{S\}$;
  \label{line:add}
  \State %%{\color{red}
    $\calC:=\calC\setminus\{S^-\in\calC\mid S^-\subsetneq S\}$
  %%}
  \label{line:exclusion}
  \EndIf
  \EndFor
  \EndFor
\EndIf;
\State Output all subsets in $\mathcal{C}$
\end{algorithmic}
\end{algorithm}

\begin{thm}
  For an instance $\calI$,
  \algref{enummaxsol} enumerates all maximal solutions. 
\end{thm}
\begin{proof}
  We show the correctness of the algorithm by induction on $k$.
  If $k=1$, then every hyperedge in $\calE$ consists of precisely
  one vertex in $V$. The unique maximal solution is the
  set $S$ of vertices that are contained in hyperedges
  and hence $|S|=|\calE|$ holds. 
  We see that \algref{enummaxsol} outputs $S$ if $k=1$ and $|S|\ge\theta_1$.

  Suppose $k>1$ and that \algref{enummaxsol} works correctly for $k-1$.
  We show that every subset $S$ in $\calC$
  is a solution (which may not be maximal) at any time in the execution of the algorithm;
  $S$ is constructed by taking the union $S=K\cup\hat{H}$ in \lineref{union}.
  The set $K$ is a subset of $V_k$ (\lineref{K})
  and also a subset of a maximal solution $B$ to $\calJ$
  that is generated in \lineref{bipartite}.
  Then $|K|\ge\theta_k$ holds. 
  Also, $\hat{H}$ is a solution to $\calI'$ in \lineref{subproblem}
  that satisfies $|\hat{H}\cap V_i|\ge\theta_i$, $i=1,2,\dots,k-1$.
  One readily sees that
  $\Pi_k(S)\subseteq\calE$
  %% $\prod_{i=1}^k(S\cap V_i)\subseteq\calE$
  holds and hence $S$ is a solution. 

  We show that any maximal solution $S$ to $\calI$
  belongs to $\calC$ when the algorithm terminates.
  By \lemref{bipartite}, there is a maximal solution $B$ to $\calJ$
  such that $S_k=B\cap V_k$,
  where $B$ is exactly generated in \lineref{bipartite}.
  Let us denote $\hat{H}:=S\setminus S_k$. 
  By \lemref{maximal}(ii),
  $\hat{H}$ is a maximal solution to $\calI'$,
  and by the induction assumption,
  the recursive call in \lineref{recursive} exactly generates
  $\hat{H}$.
  Then $S=S_k\cup\hat{H}$ is added to $\calC$ in \lineref{add}
  since it is inclusion-wise maximal and hence no
  maximal solution $S^+\supsetneq S$ to $\calI$ exists.  
  Once $S$ is added to $\calC$,
  it is not excluded from $\calC$ in \lineref{exclusion}. 

  A non-maximal solution cannot belong to $\calC$
  at the end of the algorithm
  since all maximal solutions are contained in $\calC$
  and any non-maximal solution
  has not been included to $\calC$ by \lineref{if}
  or has been excluded from $\calC$ in \lineref{exclusion}. 
  Then $\calC$ is the set of all maximal solutions
  when the algorithm terminates. 
\end{proof}

Let us make remarks on our Python implementation of the algorithm.
In \lineref{bipartite}, we enumerate
bicliques in $\sfB_\sfH$ by the algorithm of \cite{Zhang:2014aa}. 
We use data structure {\tt set}
to realize the family $\calC$ of candidate solutions,
which is essentially a hash table. 

% LocalWords:  hypergraph hyperedges partite hyperedge maximality NP
% LocalWords:  EnumMaxSol enummaxsol celltype biclique bicliques

\subsection*{Generalized linear model analysis}
A GLM was used to evaluate the contribution of discrete variables to gene expression levels, which served as the objective variable. Discrete variables, such as sex, tissue, and cell-type, were used as explanatory variables. The regression coefficients and P-values for the discrete variables were calculated using the ``glm'' function in R with default settings. The P-values for each discrete variable were calculated using Type II ANOVA, implemented with the ``Anova'' function from the R package ``car'' (version 3.1.3) \cite{car}.

\subsection*{Gene Ontology analysis}
Gene Ontology analysis was performed for the biological interpretation of the results derived from the omics analysis performed using the proposed method.
Gene Ontology analysis was performed using the ``WebGestaltR'' package (version 0.4.6) in R \cite{liao2019webgestalt}. The ``WebGestaltR'' function within this package was used for the analysis, and Over-Representation Analysis (ORA) was specified as the method. The analysis focused on the GO category ``Biological Process'', using the database setting ``geneontology\_Biological\_Process\_noRedundant''. 

\subsection*{Data visualization}
The bipartite graph was visualized using the ``igraph'' (version 1.6.0) and ``multigraph'' (version 0.99) packages in R \cite{igraph, multigraph}.

\section*{Acknowledgement}
 We would like to thank FORTE Inc. (https://www.forte-science.co.jp/) for proofreading the manuscript.
 
\section*{Declarations}
\subsection*{Funding}
This research has been supported by the Kayamori Foundation of Informational Science Advancement.

\subsection*{Conflict of interest}
The authors have no conflict of interest to declare.

\subsection*{Ethics approval and consent to participate}
Not applicable.

\subsection*{Consent for publication}
Not applicable.

\subsection*{Author contribution}
DO designed the research. DO, KH and JZ performed the research.
JZ, KS and YN contributed to algorithm development and program coding.
DO analyzed the data. DO and KH wrote the paper.

\subsection*{Code availability}
The R code used for the simulation and data analysis is available at https://github.com/DaigoOkada/scRNAseq-biclique.
The developed software is available at https://github.com/ku-dml/k-partite-hypergraph.

\subsection*{Software Availability and Requirements}
\begin{itemize}
    \item Project name: k-partite-hypergraph
    \item Project home page: https://github.com/ku-dml/k-partite-hypergraph
    \item Operating system(s): Platform independent
    \item Programming language: Python
    \item Other requirements: Python 3.9 or higher, pandas $\geq$ 1.4.2
    \item License: MIT License
    \item Any restrictions to use by non-academics: License needed
\end{itemize}

%arxivでは消す
%\nolinenumbers

\clearpage
%This is where your bibliography is generated. Make sure that your .bib file is actually called library.bib
\bibliography{library}

%This defines the bibliographies style. Search online for a list of available styles.
\bibliographystyle{abbrv}

\section*{Supporting Information}
Legends to Supplementary Files

{\bf Supplementary File 1}. All combinations of tissues and cell-types in the FACS and Droplet datasets.

{\bf Supplementary File 2}. All maximal solutions identified from the combinations of tissues and cell-types in the FACS dataset.

{\bf Supplementary File 3}. All maximal solutions identified from the combinations of individuals, tissues, and cell-types in the FACS dataset.

{\bf Supplementary File 4}. P-values for tissue and cell-type effects for all genes from the statistical analysis of 24 sub-datasets derived from the FACS dataset.

{\bf Supplementary File 5}. Complete results of the Gene Ontology (GO) analysis for genes affected by the tissue environment in the FACS dataset. The table was generated using the ``WebGestaltR'' package in R.

{\bf Supplementary File 6}. P-values for tissue and cell-type effects for all genes from the statistical analysis of one sub-dataset derived from the Droplet dataset.

{\bf Supplementary File 7}. Complete results of the Gene Ontology (GO) analysis for genes significantly affected by the tissue environment in the sub-dataset of the Droplet data (limb muscle vs. mammary gland). The table was generated using the ``WebGestaltR'' package in R.

{\bf Supplementary File 8}. Directed acyclic graphs (DAGs) constructed for all 54 genes identified in this analysis.

{\bf Supplementary File 9}. Complete results of the regression analysis from the aging study (BAT vs. SCAT). The table shows regression coefficients, P-values, and false discovery rate (FDR) values for "Young" category in each tissue.

{\bf Supplementary File 10} Complete results of the regression analysis from the aging study (MAT vs. SCAT). The table shows regression coefficients, P-values, and false discovery rate (FDR) values for the "Young" category in each tissue.

{\bf Supplementary File 11} Complete results of the Gene Ontology (GO) analysis for genes with expression changes in opposite directions between BAT and SCAT. The table was generated using the ``WebGestaltR'' package in R.

\end{document}